\documentclass[conference]{IEEEtran}
\IEEEoverridecommandlockouts
\usepackage{cite}
\usepackage{comment}
\usepackage{amsmath,amssymb,amsfonts}
\usepackage{algorithmic}
\usepackage{graphicx}
\usepackage{textcomp}
\usepackage{xcolor}
\def\BibTeX{{\rm B\kern-.05em{\sc i\kern-.025em b}\kern-.08em
    T\kern-.1667em\lower.7ex\hbox{E}\kern-.125emX}}
\begin{document}

\title{Privacy Protection of Grid Users Data with Blockchain and Adversarial Machine Learning\\

}

\author{\IEEEauthorblockN{Ibrahim Yilmaz\IEEEauthorrefmark{1},
Kavish Kapoor\IEEEauthorrefmark{2},
Ambareen Siraj\IEEEauthorrefmark{3},
Mahmoud Abouyoussef\IEEEauthorrefmark{4}
}
\IEEEauthorblockA{\textit{Department of Computer Science} \\
\textit{Tennessee Technological University}\\
Cookeville, USA 
\\iyilmaz42\IEEEauthorrefmark{1},
asiraj\IEEEauthorrefmark{3},
mabouyous42\IEEEauthorrefmark{4}@tntech.edu\\ kavishkapoor97\IEEEauthorrefmark{2}@gmail.com }}

\maketitle

\begin{abstract}
Utilities around the world are reported to invest a total of around \$30 billion over the next few years for installation of more than 300 million smart meters, replacing traditional analog meters \cite{info}. By mid-decade, with full country wide deployment, there will be almost 1.3 billion smart meters in place  \cite{info}. Collection of fine-grained energy usage data by these smart meters provides numerous advantages such as energy savings for customers with use of demand optimization, a billing system of higher accuracy with dynamic pricing programs, bidirectional information exchange ability between end-users for better consumer-operator interaction, and so on. However, all these perks associated with fine-grained energy usage data collection threaten the privacy of users. With this technology, customers’ personal data such as sleeping cycle, number of occupants, and even type and number of appliances stream into the hands of the utility companies  and can be subject to misuse. This research paper addresses privacy violation of consumers’ energy usage data collected from smart meters and provides a novel solution for the privacy protection while allowing benefits of energy data analytics. First, we demonstrate the successful application of occupancy detection attacks using a deep neural network method that yields high accuracy results. We then introduce Adversarial Machine Learning Occupancy Detection Avoidance with Blockchain (AMLODA-B) framework as a counter-attack by deploying an algorithm based on the Long Short Term Memory (LSTM) model into the standardized smart metering infrastructure to prevent leakage of consumer’s personal information. Our privacy-aware approach protects consumers' privacy without compromising the correctness of billing and preserves operational efficiency without use of authoritative intermediaries.
\end{abstract}


\section{Introduction}

There are more smart meters being deployed around the world today than conventional electromechanical/analog ones. For instance, according to \cite{smart_meter}, 107 million smart meters were planned to be installed in the USA by the end of 2020. With traditional analog meters, utility workers manually report consumers’ energy consumption usage on monthly basis. However, digital smart electric meters enable automatic measurement of energy usage and reporting in a timely and controlled fashion. Such high frequency and time-based consumption data offer a variety of benefits. For example, it helps customers to consume less energy when demand on the smart power grid is very high. Also, this technology allows the grid users to generate their own power, and even sell unused energy back to the providers. Furthermore, on the utility side, it tracks the power flows continuously to prevent blackouts, reducing waste of energy, and so on.

However, all these perks associated with the collection of fine-grained energy usage data threaten the privacy of users. Highly granular smart meter data inevitably pour private information of inhabitants into the hands of the utility companies. This confidential information can be sold by energy companies to interested third parties or government agencies for subsidiary revenue or it can be stolen by adversaries through unauthorized access or security breaches. Such personal information extracted from users’ energy consumption data can disclose details of customers’ daily life and later can be used for malicious purposes. For example, it can be used to infer when a homeowner is at home. Therefore, a burglar can break into a vacation home when the owner is absent. Even the presence of a home security alarm system or other appliances can be identified by analyzing the power signature. The same can be used by burglars to pre-identify and plan the home invasion.

The primary goal of this research is to propose a mechanism to minimize the trade-off between privacy protection and data-utility. We introduce Adversarial Machine Learning Occupancy Detection Avoidance with Blockchain (AMLODA-B) framework as a counter-attack against adversarial occupancy detection by deploying an algorithm based on the Long Short Term Memory (LSTM) model into the standardized smart metering infrastructure to prevent leakage of consumer’s personal information. Our privacy-aware approach ensures consumer privacy without compromising the correctness of billing along with operational efficiency and no requirement of authoritative intermediaries.

Initially, the proposed AMLODA model uses historical data to capture subtle and functional patterns of occupancy through a training process and create a customized pre-trained model. Then, during testing, the proposed algorithm automatically generate noisy samples that can conceal time-of-use information of consumers. In our earlier work \cite{siraj2020avoiding}, we offered that the pre-trained models can be developed with help of a trusted third party in the initialization phase. For this, a group of consumers with diverse energy load profiles can be recruited by a trusted third party through an incentivized volunteer program, who would provide access to their energy usage profiles used to pre-train the models. However, there are two potential challenges with this approach: 1) users may not want to volunteer to share their consumption information due to privacy reasons; 2) the third party’s presence in the initialization phase poses additional privacy and security challenges. In addressing these issues, we present the new AMLODA-B framework with aid of blockchain technology, which allows us to gather the information of all grid users in a privacy-preserving manner while eliminating the need of the trusted third party. The proposed scheme helps utility companies to create the pre-trained models for users without releasing their personal information. Furthermore, we evaluate the AMLODA-B model’s performance under a black-box setting by implementing and exercising both an attack and a defense scenario as part of this study.

Our work has the following contributions:
\begin{itemize}
\item We demonstrate the viability of occupancy detection attack using a machine learning technique based on the neural network model.  
\item We use blockchain technology to securely collect users’ consumption data to develop pre-trained energy profile models that are used by the AMLODA-B framework. The user governs their energy usage data for the purpose of fulfilling their privacy needs with this proposed scheme. 
\item To demonstrate AMLODA-B model’s effectiveness, we evaluate the model’s performance under a black-box scenario without knowledge of the occupancy attack models’ inner workings.

\end{itemize}

The rest of this paper is organized as follows: The literature review is discussed in Section \ref{related}. The preliminaries relevant to our study is reviewed in Section \ref{background}. We address the system models and design goals in Section \ref{system}. Section \ref{implementation} describes the implementation of our model along with blockchain. We evaluate the model's performance in Section \ref{result}. In conclusion, we finalize the paper in section \ref{conclusion}.

\section{Related Work}
\label{related}
Researchers have been working on privacy preserving solutions for smart meter users over the past few years. One proposed solution is aggregation of users’ metering data via a trusted third party (TTP) \cite{erkin2013privacy}, \cite{ford2017secure}. In this approach, the TTP, who provides trusted billing service for customers as aggregator, also protects their confidential information from utility companies as a trade secret. However, such solutions work under the assumption that the TTP is fully trusted by all users and behaves accordingly. This makes the proposed models impractical as trust cannot always be guaranteed. To overcome the above-mentioned limitation of privacy-preserving data aggregation models, M{\'a}rmol, Sorge, Petrlic, Ugus, Westhoff, and P{\'e}rez  \cite{marmol2013privacy} proposed a privacy-enhanced architecture to keep personal information of users secret from both the utility provider and the TTP itself. This scheme allows each individual smart meter to encrypt its electricity usage with its own key and a group key using a ring-based topology. Once the TTP acquires the aggregated measured data in an encrypted form, it can decrypt the total electricity consumption of the consumers. However, this scheme is not efficient for latency-critical smart grid services because of using homomorphic encryption. Furthermore, this approach does not support non-repudiation because of using asymmetric encryption with shared secret. In addition to this, reconfiguration is needed every time a node/smart meter joins or leaves the network. 

Another alternative solution is anonymizing smart meters from any untrusted entities including the utility provider \cite {diao2014privacy}, \cite{chu2013privacy}. However, such solution is still not sufficient to protect users’ private information because anonymized data still discloses important auxiliary information about a household such as the number of occupants, number of school-age children in a home. Such information can possibly aid to infer a user’s identity.

Another proposed method offers to use additional hardware that is referred to as battery-based load hiding to mask users' sensitive information \cite{tan2013increasing}, \cite{liu2017information}. Direct manipulation of the real energy consumption can be achieved by charging or discharging this external battery and provide privacy protection for customer. However, battery longevity and cost make this privacy-preserving solution questionable for adaptation. Additionally, the success of this solution is largely storage capacity-dependent and it does not support dynamic pricing.  

\section{Preliminaries}
\label{background}
In this section, the preliminaries related to our research is presented.

\subsection{Blockchain Technology:}
Blockchain is a decentralized and distributed record for digital assets \cite{drescher2017blockchain}. The basic building block in a blockchain is known as a “block”. A “genesis” block acts as a starting point for the blockchain to which other blocks are added. Blocks in a blockchain cannot be modified once added. Chains are built as records for the data to be stored are added in the digital ledger. Each block is linked to the chain using cryptography by containing a cryptographic hash of the previous block, a timestamp, and transaction data. The hash for each block is determined using a hash function. A hash function is a function that takes input data of any size and produces hash of a fixed size, which is a unique representation of data that can be used to identify the input. Some commonly used hash functions are Secure Hashing Algorithm (SHA-2 and SHA-3), RACE Integrity Primitives Evaluation Message Digest (RIPEMD), Message Digest Algorithm 5 (MD5), BLAKE2, etc \cite{sobti2012cryptographic}.

A blockchain can be used to validate new input blocks and avoid overwriting of old blocks using some proposed schemes such as proof of work or proof of stake \cite{bentov2014proof}. Proof of work is a form of cryptographic zero-knowledge proof wherein a block is added by a certain individual or party only when it is proved that a certain amount of computational effort has been expended for some purpose. Our proposed system makes use of the proof of work algorithm to validate and add new blocks. One common example of the proof of work algorithm is the Hashcash algorithm, which is most commonly used in bitcoin and some other cryptocurrencies. The process of placing unconfirmed transactions in a block and computing proof of work is known as block mining. Only after this validating step using proof of work is executed, a block is successfully added to the blockchain.

\subsection{Machine Learning models}

\begin{itemize}
\item \textbf{Long Short-Term Memory (LSTM) Model:} The LSTM, a specialized Recurrent Neural Network (RNN), can work with single data points as well as sequences of data. LSTM model consists of three gates named input, output, and forget gates \cite {woodbridge2016predicting}. These three cells regulate the flow of information through the LSTM unit and determine which part of the information is important to be kept or to be discarded.

\item \textbf{Deep Neural Network (DNN):} DNN consist of multiple layers called input, hidden and output layers. Each layer is made of nodes that are connected to each other \cite{yilmaz2019expansion}. The model learns the most important features in high dimensional data during backward and forward propagation using a gradient descent algorithm \cite {anderson1995introduction}. Activation functions are applied after each of the following layers to make the model capture the nonlinear relationship between input and output in solving difficult tasks \cite{yilmaz2020practical}.

\end{itemize}

\section{System Models and Design Goals}
\label{system}
In this section, we describe system entities along with the attack and defense models we used to demonstrate the use and applicability of our approach. The features of the AMOLDA-B framework are also articulated.

\subsection{System Entities}

\begin{itemize}
\item \textbf{Grid Users (GUs):} GUs have smart meters installed in their households and have agreed to use the AMLODA-B framework for their smart meter data communication. To setup the process, GUs initially share their energy usage data with the utility company using the blockchain technology associated with AMLODA-B prior to the actual privacy preserving usage data communication activation. Important to note that even at this setup stage, the utility company does not have the ability to associate the smart metering data with the generating smart meter/customer apart from collecting them for pre-trained model development and thus, maintaining the GUs’ privacy. Once the customized pre-trained model is developed and the system is ready to be turned on, a user selects the desired privacy level from different options provided by the AMLODA-B framework. To educate GU on such selection of privacy levels, the system can provide some helpful information on its implication to users’ privacy vs energy efficiency.         
\item \textbf{Utility Company (UC):} Using blockchain technology, the UC collects energy usage data from the smart meters belonging to volunteer GUs to create the pre-trained model offline needed for training of the AMOLDA-B model. Once the pre-trained models are developed for each GU, the UC sends the model information back to the smart meters through the blockchain. Since the blockchain ensures tamper-proof data communication, if a malicious GU tamper with the measured consumption readings, it will easily be detected.  

\item \textbf{Key Distribution Center (KDC):} KDC is the Trusted Third Party, who is responsible for generating public/private key pairs for each GU and UC. KDC has no access to the blockchain itself in our proposed scheme.

\end{itemize}

\subsection{Attack Model}
\begin{itemize}
\item \textbf{Attacker's Knowledge:} In this attack scenario, we assume that the attacker is able to access the GUs’ actual smart meter energy consumption data. This can happen in the real-world scenario when either the UC itself is the adversary or it colludes with an external adversary by sharing of GUs’ sensitive information for subsidiary revenue.    

\item \textbf{Attacker's Goal:} Even if the UC behaves more like an honest-but-curious adversary, meaning that it abides by the protocol rules and does not provide false information, it does attempt to learn possible private information of the grid users in an unauthorized way. In real world scenario, this private information has lot of market value with advertising and marketing agencies. For the remaining of this section, we will refer to the UC as the adversary.

\end{itemize}
\subsection{Defense model}
\begin{itemize}
\item \textbf{Defender's Knowledge:} The defender, who has best interest of the GU, has full access to GU's energy consumption data like the attacker. However, the defender has zero knowledge about the adversary’s pre-trained model. For example, which algorithm or which parameters are used in the model are not known by the defender.

\item \textbf{Defender's Goal:} The defender's aim is to create a black-box model (AMLODA-B model) to rearranges the GUs' electricity consumption data by means of noise injection. The model takes the actual consumption of the GUs and generates scramble data that the adversary receives. The adversary is unable to learn any usable information from new flow patterns.    

\textbf{Note:} The defender makes use of the transferability property of adversarial samples. This means that samples crafted to mislead the defender's model are more likely to mislead the attacker's model as well because when the defender's and attacker's models are trained with data that comes from the same distribution, both models learn similar decision boundaries.     
\end{itemize}

\subsection{AMLODA-B Features:}
The proposed AMLODA-B framework fulfills the following functional and privacy requirements. 

\textit{1) Functional Requirements:}

\begin{itemize}
\item \textbf{Correctness of billing:} The proposed method provides evidence of correctness of the billing invoice along with compatibility with dynamic pricing. Smart meter reading data is symmetrically raised and lowered by calculated noise for every two-second period such that the correctness of total energy consumption of the GU remains preserved by end the two-second intervals. Since negligible changes performed in short time periods, the calculated optimum noise does not compromise the correctness of billing and dynamic pricing.   

\item \textbf{Preserving demand management functionality:} The demand management program in the advanced metering infrastructure of the smart grid provides an opportunity for the GUs to contribute in the operation of the electric grid by reducing or shifting their electricity usage during peak periods by closely observing their energy consumption patterns. Thus, any significant changes in energy usage patterns can hinder this functionality and cause detrimental impact on the smart grid demand management system. However, this study shows that the proposed framework can provide an adequate level of privacy with negligible changes on smart meter readings and thus preserving this crucial functionality of modern meters. 
\end{itemize}

\textit{2) Privacy Requirements:}

\begin{itemize}
\item \textbf{Preserving privacy of GUs energy usage data:} AMLODA-B supports privacy-preserving mechanism to prevent disclosure of GUs private energy consumption information by integrating a machine learning model with blockchain technology. Initially, blockchain technology is used to collect energy usage data by the UC. Then, pre-trained model of customer energy consumption is developed using machine learning technique with LSTM. Finally, the proposed perturbation algorithm runs on the pre-trained model to reproduce energy consumption data in a way that ensures integrity of billing but at the same time, does not give away real consumption data revealing consumer’s energy profile. In this way, actual information is kept locally on the user side and no entity, either internal or external including the UC, is able to see access consumption readings of the GUs at any reporting period.

It is worthwhile to point out that both blockchain and machine learning techniques require a long time to complete execution. However, in our approach, the pre-trained model is developed offline before used with the perturbation algorithm in real-time, which leads to lower computational complexity. This causes no or negligible transmission delay in time-critical traffic.

\end{itemize}

\section{Model Implementation}
\label{implementation}

\subsection{Dataset}
\label{dataset}
Electricity Consumption and Occupancy (ECO) dataset has been provided by ETH Zurich for the community in an attempt to motivate researchers to contribute in the field of smart meter consumers' privacy protection \cite{dataset}. The experimental dataset includes power consumption readings every seconds in addition to ground-truth occupancy information. Data was collected during the period between June 2012 to January 2013 by monitoring 5 different residences in Switzerland. 
\subsection{Occupancy Detection Attack Implementation}

To demonstrate viability of occupancy detection attacks, we design one such attack based on a DNN in Python programming language using the Keras library. The reasons for selecting a DNN model for this task are that a DNN model usually performs well with a large dataset and we want to demonstrate the transferability property of the crafted samples by the AMLODA-B model over different types of machine learning models. This enables better understanding of the effectiveness of the AMLODA-B framework under a realistic scenario.

The experimental dataset is divided into two parts as training and test where 80\% of the dataset is used to train the model and the remaining is used for evaluation of the model’s performance. The DNN model consists of four hidden layers along with the input and output layers. Table \ref{table} demonstrates the detailed structure of the occupancy attack model including \textit{number of layers}, \textit{number of neurons}, and a\textit{ctivation functions}. 

\begin{table}[!ht]
\centering
\includegraphics[width=8cm]{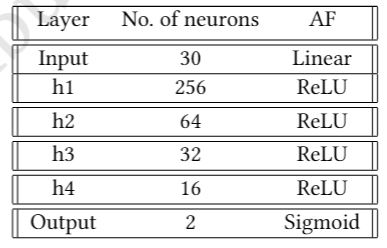}
\caption{Parameters of the Occupancy Attack Model.}
\label{table}
\end{table}

The fine-tuned parameters are discovered by using a \textit{learning rate} of 0.001, \textit{epoch number} of 30000, \textit{adam optimization algorithm} \cite{kingma2014adam} along with \textit{binary cross entropy} loss function \cite{murphy2012machine}.

\begin{table*}[!ht]
\centering
\includegraphics[width=14cm]{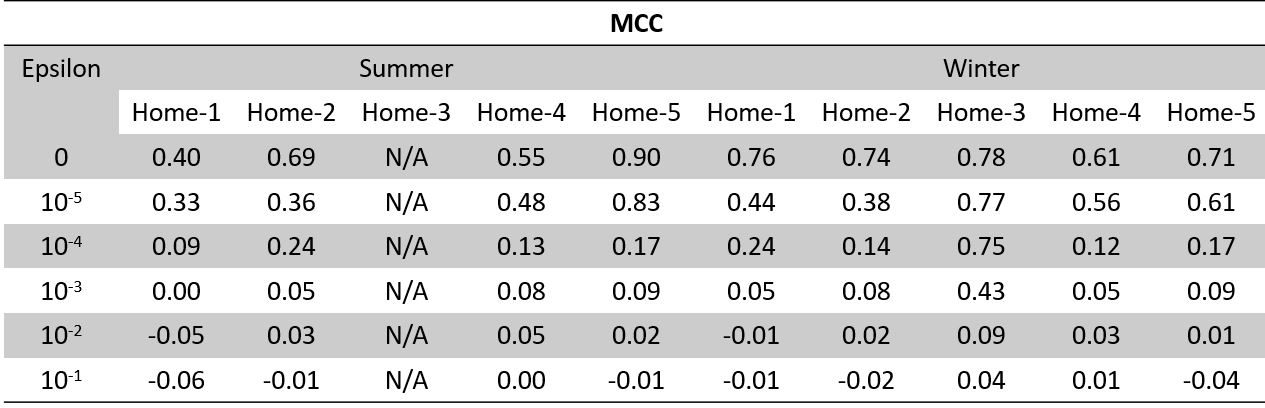}
\caption{MCC vs. penetration coefﬁcient for ﬁve houses during summer and winter periods.}
\label{mcc}
\end{table*}

\subsection{AMLODA-B Model }

We implement the AMLODA-B model based on the LSTM with Python programming language using Pytorch library. Learning usable information for occupancy detection requires ability to learn long-term dependencies. In such circumstances, the LSTM model is a good choice to capture complex time-series dynamics by avoiding gradient vanishing problem \cite {hochreiter1998vanishing}. Initially, the model learns the characteristic behavior of users’ consumption through their historical data. Then, small indiscernible perturbations are generated in an adversarial way, a technique inspired by \cite{goodfellow2014explaining}. These perturbations are added to the actual meter reading in selected time intervals to mask real electricity consumption patterns for avoiding occupancy detection. While adding a large amount of noise increases the user’s privacy to a large extent, such modification causes substantial changes in load patterns which might have a negative impact on the demand management system. Therefore, the proposed AMLODA-B model allows users to select appropriate noise levels (including no noise) that best meets their need of privacy-preservation and energy efficiency. The mathematical foundation of the proposed scheme along with the model’s parameters were discussed in detail in \cite{siraj2020avoiding}.

\subsection{Blockchain details}
The blockchain plays a pivotal role in masking the grid user’s identity for pre-train model development and afterwards, supplying the utility company with the perturbed data from household smart readings. The blockchain is simulated and implemented using python 3.7 running on a standard desktop machine with an Intel Core i7-7700HQ Central Processing Unit (CPU) operating at 2.80GHz and 16GB of Random Access Memory (RAM). Each new block that is added to the blockchain has the following structure:  

\{'index': index for the current block in the blockchain, 'transactions': list of the energy readings data and occupancy data from one house, 'timestamp': timestamp for the current block, 'previous\_hash': hash of the previous block, 'nonce':nonce value, 'hash':hash value\}.

The index represents the index assigned to the current block in the blockchain. The key 'transactions', is a list of all the energy readings from various connection plugs in a household. The transactions list also contains the information for household occupancy for each second over a day partitioned
into 2 sessions for 'summer' and 'winter'. The timestamp key refers to the time since the epoch of the creation of the block that is to be added to the blockchain. The 'previous\_hash' key refers to the hash of the previous block and is used as a pointer to the previous block in the blockchain. The next key is 'nonce' , which is a number that we can change until we get a hash that satisfies constraints thus serves as proof that some computation has been performed. It is used in the \textit{hashcash} algorithm \cite {back2002hashcash} that is implemented in the blockchain as a proof of work algorithm. Lastly, the ’hash’ key represents the hash of the current block that has to be added to the blockchain. SHA-256 is used as the hashing function in the blockchain’s implementation.

\section {Evaluation and Results}
\label{result}
\subsection{Blockchain Performance}

The blockchain was used to simulate the collection of real-time energy usage data from 5 households. Individual blocks comprised of the energy usage data from one house on average, took 102.8239 seconds to get validated and added to the blockchain. The total time taken for the addition of all the energy usage data from 5 houses took 514.1262 seconds. The whole blockchain computation used a space of 2.45 GB.

\subsection{AMLODA-B Counter Attack Model Performance}

To evaluate the AMLODA-B model’s performance against a realistic occupancy detection attack based on a deep neural
network, we first monitored the neural network model’s performance with various data perturbations (epsilon values) using Matthews Correlation Coefficient (MCC). MCC ensures reliable assessment when a dataset is highly imbalanced such as in this case \cite{yilmaz2020addressing}, \cite{chicco2020advantages},\cite{yilmaz2020improving}. MCC indicates model’s performance in a range of -1 to 1 \cite{kantardjieff2003matthews}, where -1 signifies completely incorrect prediction, 0 denotes random guessing and 1 indicates perfect prediction. MCC value converging to 0 is best for hiding consumers’ sensitive information  since in such instance, inference is no better than random guessing.

Initially the epsilon value was set to zero to show original data without any manipulation. Then, the calculated noise, obtained by the AMLODA-B model using different small epsilon values, was added to data to observe how the perturbed data samples impact the performance of the attack model for the five houses over summer and winter periods. As seen in Table \ref{mcc}, the attack model shows high detection rate for occupancy with high MCC values without use of the proposed scheme. With the proposed AMLODA-B model in place, the epsilon value is scaled up until the MCC value of the attack model converges to zero. Since perturbed data is no longer representative of the distribution of the users’ consumption readings, therefore, the model now mispredicts them. The table also shows that the AMLODA-B model obscures consumers’ privacy in most cases with epsilon value set to $10^{-4}$, which makes an insignificant change over actual load pattern \cite{siraj2020avoiding} without causing any inefficiency in smart grid operations. It should be noted that this value is not able to sufficiently protect sensitive personal information of grid users for Home-3 in the winter season. If the value of epsilon is increased higher, the MCC score converges to zero signifying that the DNN-based occupancy detection attack becomes equivalent to random guess. It is worth noting that Home-3 information for the summer season is not made available to the public and hence could not be used in the experiment.


\section{Conclusion and Future Work}
\label{conclusion}

In this paper, we highlight privacy leakage problem of smart meter users by implementing an occupancy detection attack based on a DNN model. Our attack model’s performance achieves high accuracy in detecting occupancy. To preserve privacy and counter occupancy detection, we presented the AMLODA-B framework using machine learning and blockchain technology. This novel scheme collects high-frequency metering data securely through a blockchain algorithm that allows to develop pre-train model without giving away identity of consumers to utility companies. Then, during actual energy usage, it hides consumers private information in real-time through crafted noise computed from observation in the pre-trained model with the aid of LSTM. While the AMLODA-B model offers data privacy service, it does not hinder the utility companies to use smart metering data for computation of consumer’s bills, demand response management, and so on. Since the proposed scheme does not require any intermediary or any additional hardware components in addition to computationally intensive software, it makes the adoption of this privacy-friendly model logistically realistic and economically feasible. 

As a future direction, we are further analyzing the load profile smart metering data to find the best perturbation amount to meet optimal productivity between privacy protection and energy efficiency. Furthermore, we will evaluate the efficiency of the proposed scheme by extending its functionality to other benefits of the smart grid such as energy forecasting and electricity theft detection.   

\section*{Acknowledgment}
This work is funded by CESR (Center for Energy Systems Research) at Tennessee Technological University with resource support from the Cybersecurity Education Research and Outreach Center (CEROC).

\bibliographystyle{IEEEtran}
\bibliography{References}
\end{document}